\documentclass[aps,prl,10pt,twocolumn,showpacs,superscriptaddress]{revtex4-2}
\usepackage{graphicx}
\usepackage{multirow}
\usepackage[dvipsnames]{xcolor}
\usepackage[normalem]{ulem}
\usepackage{dcolumn}
\usepackage{amsmath}
\usepackage{amssymb}

\newcommand{\sgn}{\mathop{\mathrm{sgn}}}

\usepackage{hyperref}

\hypersetup{
    hidelinks,
    colorlinks=true,
    linkcolor=blue,
    filecolor=blue,      
    urlcolor=blue,
    citecolor=blue
}

\begin{document}

\title{Electronic Polarization Effects in Core-Level Spectroscopy}

\author{Iskander~Mukatayev}
\affiliation{Universit\'e Grenoble Alpes, CEA, Leti, F-38000, Grenoble, France}

\author{Gabriele~D'Avino}
\affiliation{Universit\'e Grenoble Alpes, F-38000 Grenoble, France}
\affiliation{CNRS, Institut N\'eel, F-38042 Grenoble, France}

\author{Beno\^it~Skl\'enard}
\affiliation{Universit\'e Grenoble Alpes, CEA, Leti, F-38000, Grenoble, France}
\affiliation{European Theoretical Spectroscopy Facility (ETSF), F-38000 Grenoble, France}

\author{Valerio~Olevano}
\email{valerio.olevano@neel.cnrs.fr}
\affiliation{Universit\'e Grenoble Alpes, F-38000 Grenoble, France}
\affiliation{CNRS, Institut N\'eel, F-38042 Grenoble, France}
\affiliation{European Theoretical Spectroscopy Facility (ETSF), F-38000 Grenoble, France}

\author{Jing~Li}
\email{jing.li@cea.fr}
\affiliation{Universit\'e Grenoble Alpes, CEA, Leti, F-38000, Grenoble, France}
\affiliation{European Theoretical Spectroscopy Facility (ETSF), F-38000 Grenoble, France}
\date{\today}

\begin{abstract}
In X-ray photoelectron spectroscopy (XPS), the injected hole interacts with the electronic polarization cloud induced by the hole itself, ultimately resulting in a lower binding energy.
Such polarization effect can shift the core-level energy by more than 1~eV, as shown here by  embedded many-body perturbation theory for the paradigmatic case of noble gas clusters made of Ar, Kr, or Xe.
The polarization energy is almost identical for the different core-orbitals of a given atom, but it strongly depends on the position of the ionized atom in the cluster.
An analytical formula is derived from classical continuum electrostatics, providing an effective and accurate description of polarization effects, which permits to achieve an excellent agreement with available experiments on noble gas clusters at a modest computational cost.
Electronic polarization provides a crucial contribution to core levels absolute energies and chemical shifts.
\end{abstract}

\maketitle

\paragraph*{Introduction. ---}
X-ray photoelectron spectroscopy (XPS) provides the binding energy of {deep} electronic levels by measuring the kinetic energy of electrons ejected by X-ray photons \cite{Siegbahn, Carlson}. 
The specific pattern of core-level binding energies (BE) grants access to the chemical composition of the studied system. 
In addition, it provides structural information, such as chemical bondings and the local environment, through the \emph{chemical shift} of the core-level BEs with respect to a reference (isolated atom or molecule). 
This requires established knowledge associating a given core-level shift with precise chemical bonding. 
While such knowledge might be acquired experimentally, as is the case of carbon for which the chemical shifts for different chemical environments are tabulated, a more promising way is to develop an accurate method to simulate core-level shift for any species in any environment \cite{Aarva_2019}.

Starting from the seminal work of Bagus \cite{Bagus_1965}, a general method to calculate core-electron BEs consists in evaluating the total energy difference between the initial, neutral ground state and the final state in the photoelectron process, that is, a charged excited state with a core hole \cite{Lindgren_2004, Vines_2018}. 
This method, called $\Delta$SCF, can be applied relying on any approach providing the total energy, from the simplest Hartree-Fock (HF) method and density-functional theory (DFT) to more advanced post-HF approaches like coupled-cluster and configuration interaction. 
The recent introduction of the $GW$ Green's function-based many-body perturbation theory to XPS simulation provides an alternative way for evaluating the core-level BE by a single calculation of the quasi-particle energy in the neutral system  \cite{Mukatayev_2022, Mejia-Rodriguez_2021, Li_Golze_2022, Galleni_2022, Golze_2020, Mukatayev_2023}.
Moreover, recently developed analytic-continuation approaches provided robust and accurate frameworks to speed up calculations of core-electron BEs. \cite{Duchemin_2020,Panades_2023}.
With the $GW$ method, the computed XPS peaks of noble gas atoms, from He to Rn, with BE up to $100$ keV, are comparable to experiments, typically with an error below $1\%$ \cite{Mukatayev_2022}.
The accuracy is also validated in a set of small molecules containing light elements, such as C, N, O, and F atoms \cite{Mejia-Rodriguez_2021, Li_Golze_2022, Golze_2020, Mukatayev_2022}.
The $GW$ formalism has been recently applied to compute core levels of organic polymers, using an effective additive approach in which long chains were partitioned into isolated monomers \cite{Galleni_2022}.

Accurately determining core levels BEs may not be necessary to correctly capture core-level shifts.
Previous studies demonstrate that the chemical shift is well captured by the change of single-particle energy in DFT calculations \cite{Pasquarello_1996, Guittet_2001, Artyushkova_2013, Mukatayev_2023, Taucher_2016}.
By decomposing the chemical shift into different terms in the Hamiltonian, recent work has shown that the main effect of the core-level shift indeed originates from an \emph{electrostatic} contribution, \cite{Mukatayev_2023} namely from the electrostatic potential generated by the charge density characterizing the local environment of the ionized atom in the neutral ground state of the system.
Dynamic \emph{electronic polarization} (i.e., screening) effects act on top of that as a reaction to the system charging (electron removal) in the photoemission process.
These are due to the interaction between the {electronic} polarization cloud induced by the core hole and the hole itself.

The polarization part is not captured in single-particle HF or DFT approaches and requires more sophisticated methods, such as $\Delta$SCF or many-body theory.
The screening effect significantly affects the absolute BEs of core levels, but it usually has little impact on energy differences between atoms, i.e., chemical shifts \cite{Mukatayev_2023}. 
However, one has to pay attention to distinguish between screening effects taking place within an isolated finite system (e.g., atom or molecule) and those for the same system in the condensed phase, where also the dielectric environment contributes to screen charged excitation.
The interplay between short and long-range screening phenomena might be rather complex, possibly leading to core-level shifts dominated by polarization effects, as discussed in the following.

In this Letter, we use embedded many-body perturbation theory to systematically investigate the polarization effect on core-level shifts.
We focus on noble gas clusters, an ideal system for this purpose that has been extensively studied experimentally with XPS \cite{Citrin_1974, Tchaplyguine, Lundwall_2006}.
Indeed, these systems are an almost perfect realization of van der Waals aggregates of neutral atoms, hence featuring negligible microscopic electrostatic fields. 
Therefore, different {peak energies} in experimental XPS spectra must be ascribed exclusively to the change in polarization energy due to a different environment, i.e., the different position of atoms, being them fully in bulk or at the surface, passing through all possible intermediate embedding conditions.
With the embedded many-body approach,  the dependence of polarization energy on crystallographic orientations and excited orbitals is examined. 
Eventually, simulated XPS spectra of noble gas clusters are compared with experiments. 

\paragraph*{Methods. ---}
The core-electron excitation of a noble gas atom in a cluster is computed within an embedded many-body perturbation theory implemented in a hybrid quantum/classical (QM/MM) approach \cite{Li_Avino_16,Li_Avino_2018}. 
The $GW$ method models the QM part by constructing the one-body Green's function
\begin{equation}
G(\mathbf{r}, \mathbf{r'};\omega) = \sum_i\frac{\psi_i(\mathbf{r}) \psi_i^*(\mathbf{r'})}{\omega-E_i+i\eta \sgn(E_i - \mu)},
\label{eqn:G}
\end{equation}
where $\psi_i$ and $E_i$ are wave functions and energies of eigenstates, $\mu$ the chemical potential, and $\eta$ a positive infinitesimal; followed by the random-phase approximation polarizability,
\begin{widetext}
\begin{equation}
\chi^0(\mathbf{r}, \mathbf{r'}, \omega) = \sum_{i, j} \left( f_j -f_i\right)\frac{\psi_i(\mathbf{r}) \psi^*_j(\mathbf{r}) 
\psi_j(\mathbf{r'}) \psi^*_i(\mathbf{r'})}{\omega - (E_i-E_j) + i\eta \sgn(E_i - E_j)},
\label{eqn:chi0}
\end{equation}
where $f_i$ is the occupation number. 
The dynamically screened Coulomb potential is
\begin{equation}
W(\mathbf{r}, \mathbf{r'}, \omega) = w(\mathbf{r}, \mathbf{r'}) +  \int d\mathbf{r_1} d\mathbf{r_2} \, w(\mathbf{r}, \mathbf{r_1}) \chi^0(\mathbf{r_1}, \mathbf{r_2},\omega) W(\mathbf{r_2}, \mathbf{r'},\omega)
\label{eqn:W}
\end{equation}
where $w$ is the effective Coulomb potential in the QM part.
In conventional $GW$ without embedding, $w$ coincides with the bare Coulomb potential $v$. Upon classical embedding, $w$ includes the screening effect from the MM part, i.e.,
\begin{equation}
w(\mathbf{r},\mathbf{r'}) = v(\mathbf{r},\mathbf{r'}) + \int d\mathbf{r_1} d\mathbf{r_2} \, v(\mathbf{r}, \mathbf{r_1}) \chi^\mathrm{MM}(\mathbf{r_1}, \mathbf{r_2}) w(\mathbf{r_2}, \mathbf{r'}),
\label{eqn:reac}
\end{equation}
\end{widetext}
where $\chi^\mathrm{MM}$ is the polarizability in the MM part. The second term in Eq.~(\ref{eqn:reac}) is the so-called reaction field.
Finally, the $GW$ self-energy is obtained
\begin{equation}
\Sigma(\mathbf{r}, \mathbf{r'}, \omega ) = i \int \frac{d \omega'}{2\pi} \, G\left(\mathbf{r}, \mathbf{r'}, \omega - \omega' \right) W(\mathbf{r}, \mathbf{r'}, \omega')
\label{eqn:Sigma}
\end{equation}
At first order in perturbation theory with respect to HF, the $GW$ quasiparticle energy is the solution of the following equation

\begin{equation}
E_i^{GW} = E_i^\mathrm{HF} + \langle \phi_i^\mathrm{HF}|\Sigma(E_i^{GW}) - v_{x}| \phi_i^\mathrm{HF}\rangle,
\label{eqn:EQP}
\end{equation}
with $v_{x}$ the Fock exchange potential.
The polarization energy from the MM part is the difference between quasiparticle energies with and without embedding. 
\begin{equation}
P_i = s_i (E_i^{GW/\mathrm{MM}} - E_i^{GW}), \label{eqn:PE}
\end{equation}
where $s_i=1/-1$ for occupied/unoccupied states. 

In practice, we performed eigenvalues self-consistent $GW$ gas-phase calculation (ev$GW$) starting from HF eigenstates. By using the ev$GW$ quasi-particle energies, single-shot 
COHSEX calculations (Coulomb-hole-screened-exchange formalism, the static version of the $GW$ approximation \cite{Hedin65}), with and without the screening from the MM part, were used to evaluate the polarization energy according to Eq.~\ref{eqn:PE}. 
The error from neglecting the frequency dependence is largely canceled by subtracting two quasi-particle energies, as demonstrated recently using the fragment $GW$ approach, explicitly accounting for the MM part's dynamical response \cite{Amblard_2022}. 

The HF eigenstates used as starting point for $GW$ calculation were obtained using the x2c-TZVPPall-2c basis set \cite{Pollak_2017} with \textsc{NWChem} \cite{nwchem}.
The embedded $GW$ calculations are performed using the FIESTA package \cite{BlaseOlevano11, Jac15a, Li_Avino_16} with the Coulomb-fitting resolution of identity technique (RI-V) \cite{Duchemin_2017} and def2-universal-JKFIT auxiliary basis sets \cite{Weigend_2008}.
The MM part is described by an atomistic induced-dipole model implemented in the MESCAL package \cite{DAvino2014}. 
The isotropic polarizability of noble gas atoms is taken from experiments \cite{Orcutt1967, Miller1978} and was set to $1.641$, $2.484$, and $4.044$~\AA$^{3}$ for Ar, Kr, and Xe, which {arrange in a} face-centered cubic structure with lattice {constants $a$ of } $5.25$, $5.59$, and $6.13$~\AA, respectively \cite{Henshaw_1958, Keesom1930, Sears_Klug_1962}.
The same structure has been used to build bulk and semi-infinite crystals and spherical nanoparticles of given radius $R_s$. The latter set to $31.7$~\AA, $33.6$~\AA\ (3000 atoms both) and $38.4$~\AA\ (3500 atoms)
for Ar, Kr, and Xe, respectively.
These cluster sizes approximately correspond to those studied experimentally \cite{Tchaplyguine}.
The reaction field matrix for atoms in an infinite bulk (at a semi-infinite surface) has been obtained by extrapolating data from explicit calculations on finite-size (half) spheres \cite{Li_Avino_2018}.

All results presented herein are obtained for a QM region consisting of a single atom. $GW$/MM results for pairs of QM atoms present a core level dimer splitting $<1$~meV, well below the numerical accuracy. This demonstrates a negligible band dispersion, as expected for strongly-localized core levels.
Identical core level energies have been obtained for single atoms and pairs embedded in the MM environment, as a result of a proper compensation of
quantum and classical polarizabilities \cite{Li_Avino_16}.

\begin{figure}[t]
\includegraphics[width=\columnwidth]{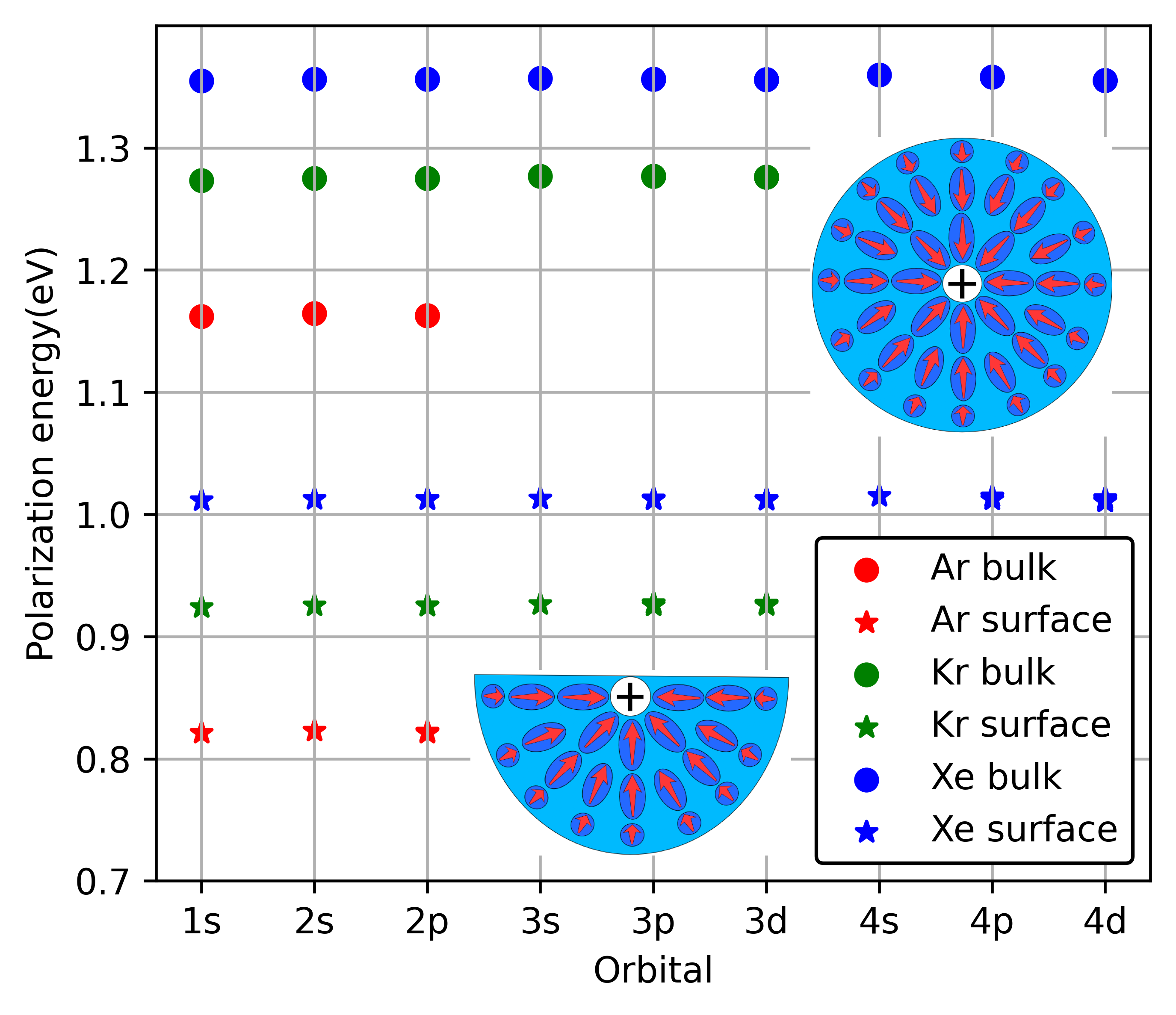}
\caption{Polarization energy induced core-level shift for all core orbitals of Ar, Kr, and Xe in bulk (sphere with infinite radius) and on a surface (semi-sphere with infinite radius).}
\label{fig:orbitals}
\end{figure}

\paragraph*{Results. ---}

\begin{figure}[t]
\includegraphics[width=\columnwidth]{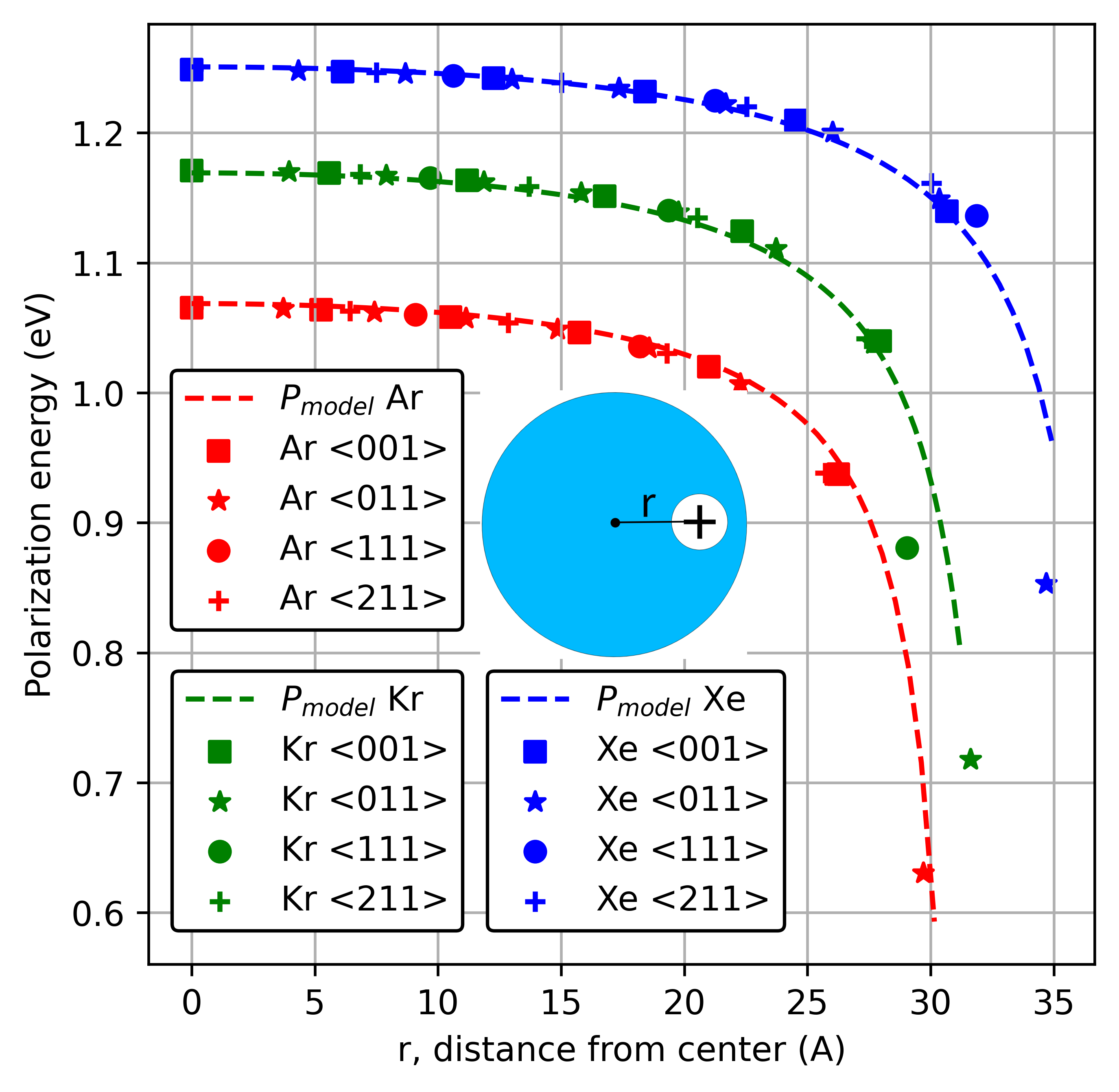}
\caption{Polarization energy of a target atom at a distance $r$ from the center of the cluster along four crystalline directions, $\langle 001 \rangle$, $\langle 011 \rangle$, $\langle 111 \rangle$, and $\langle 211 \rangle$.
The dashed lines plot the analytic formula Eq.~(\ref{eqn:analyt}).}
\label{fig:polarization}
\end{figure}

Considering a noble gas atom in its bulk solid, Fig.~\ref{fig:orbitals} shows that the polarization energy $P_i$ for a given element barely depends on the considered atomic orbital $i$.
The almost orbital-independent $P$ is about $1.16$, $1.27$, and $1.36$~eV for Ar, Kr, and Xe, respectively.
The difference in polarization energy between different elements is well captured by the Born equation, a celebrated result of classical electrostatics for the polarization energy of a point charge in a spherical cavity of a dielectric, 
\begin{equation}
P = -\frac{e^2}{8 \pi \epsilon_0 r_{c} } \left(1-\frac{1}{\epsilon_r}\right),
\label{eqn:born}
\end{equation}
where $e$ is the unit of charge, $\epsilon_0$ the vacuum dielectric permittivity, $\epsilon_r$ the relative dielectric constant, and $r_c$ the cavity radius. 
Indeed, by taking the Clausius-Mossotti dielectric constant ($\varepsilon_r=$1.70, 1.94 and 2.25 for Ar, Kr and Xe, using experimental atomic polarizability and volume) and a plausible cavity radius of $a/2$, the $GW$/MM results can be reproduced with Eq.~\ref{eqn:born} within 50~meV. This agreement allows ascribing the trend in $P$ along the series of noble-gas elements to the differences in atomic radius and polarizability.

For an atom at the surface, the polarization energy reduces by $0.2$ to $0.3$ eV with respect to the bulk, see Fig.~\ref{fig:orbitals}. However, it remains independent on the core level, about $0.82$, $0.92$, and $1.01$~eV for Ar, Kr, and Xe, respectively. The reduction of the polarization energy on the surface, as compared to the bulk, is expected since the atom at the surface feels a reduced polarizable environment around it.

The fact that the polarization energy is the same for all core levels of a given atom is a direct consequence of the Gauss theorem. 
Indeed, since the charge density of core orbitals is well localized around the nucleus, with negligible tails extending in the proximity of nearest neighbors, the field acting on the surrounding atoms is independent of the ionized level, hence determining level-independent reaction fields on the excited atom.
The reaction field of the induced dipoles is zero by symmetry at the center of the cavity (position of the QM atom nucleus), and the potential varies slowly around this point. This rationalizes why different core levels, all closely localized around the nucleus, feature nearly identical polarization energies.

In the following, we discuss finite-size clusters. 
The polarization energy of the atom at the center of a cluster of radius $R_s$ is smaller than in bulk because of the finite polarizable medium surrounding it, which provides a weaker dielectric screening.
Considering a cluster about $3$~nm in radius, the polarization energy of the central atom is $1.06$, $1.17$, and $1.25$~eV for Ar, Kr, and Xe, respectively, which are about $0.1$~eV smaller than in bulk. 
We have monitored the evolution of the core levels across the cluster by probing atoms going from the center of the cluster toward the surface.
In Fig.~\ref{fig:polarization}, we consider four crystalline directions, i.e., $\langle 001 \rangle$, $\langle 011 \rangle$, $\langle 111 \rangle$, and $\langle 211 \rangle$, and plot the polarization energy as a function of the distance from the center $r$.
The decreasing trend with $r$ reflects the fact that as probed atoms approach the surface, they become less and less embedded in the polarizable medium, hence receiving a weaker reaction field.
Interestingly, the direction dependence is very weak due to the cubic crystal symmetry: only the distance to the center matters, except near the crystal surface that breaks the bulk symmetry, resulting in a weak anisotropy.
Finally, the polarization energy of an atom at the surface of a finite spherical cluster is much smaller ($0.63$,  $0.72$, and $0.85$ eV for Ar, Kr, and Xe) than on an infinite planar surface.
This is due again to the finite size of the cluster.
Moreover, an atom at the surface of a cluster is, in fact, on a convex surface, more exposed to vacuum than to the polarizable medium.

To explain the polarization energy versus distance from the center, we derived an analytic formula.
The target atom is located at a distance $r$ from the center in a cavity. 
The polarizable medium is a sphere of radius $R_s$.
According to classical electrostatics, the polarization energy has the following form (the detailed derivation is available as Supplemental Material)
\begin{widetext}
\begin{equation}
P(r) = - \frac{e^2}{8 \pi \epsilon_0 } \left(1-\frac{1}{\epsilon_r}\right) 
\left( \frac{1}{r_c} - \frac{1}{2} \frac{R_s}{R^2_s - r^2} + \frac{1}{4r} \ln \frac{R_{s}-r}{R_{s}+r}
\right), \quad
 r\in[0,R_s-r_c].
\label{eqn:analyt}
\end{equation}
\end{widetext}
Such an equation reduces to the Born equation with $R_s \to \infty $ and $r=0$.
With a single explicit $GW$/MM calculation, e.g. for an ionized atom in an infinite bulk and dielectric constant from the Clausius-Mossotti relation, one can exploit Eq.~\ref{eqn:born} to obtain cavity radii $r_c$ of $2.55$, $2.74$, and $2.95$~\AA\ for Ar, Kr, and Xe, respectively.
With these parameters, the distance dependence is well captured by the analytic formula, see Fig.~\ref{fig:polarization}.
Discrepancies are limited to atoms at the surface, reflecting the limits of the continuum polarizable model in describing an atomistic, i.e. discrete, system.

Thus, in order to compare with XPS experiments, the polarization energy of any atom in a cluster can be quickly evaluated with a single $GW$ calculation and the analytic formula Eq.~(\ref{eqn:analyt}) instead of performing many heavy embedded $GW$ calculations.
In order to model experimental spectra, we explicitly considered the decay of the XPS signal intensity with the position of the ionized atom from the sample surface. The XPS intensity is taken proportional to $\mathrm{exp}(-|r-R_s|/\lambda)$, where $\lambda$ is the decay length (electron inelastic mean free path),  set to $8$~\AA\ for all systems.

The $GW$ BEs for atomic Ar $2p_{1/2}$, Kr $3d_{3/2}$, and Xe $4d_{3/2}$ are $-251.9$, $-94.9$, and $-69.3$ eV, in good agreement with experiments $-250.6$, $-95.0$, and $-69.5$ eV, respectively \cite{Lotz70}. To focus on the polarization effect, in the following we take the experimental peaks with the experimental spin-orbit split as $2.1$, $1.2$, and $2.0$ eV for Ar 2p, Kr 3d, and Xe 4d \cite{Lotz70}.

\begin{figure}[bt]
\includegraphics[width=\columnwidth]{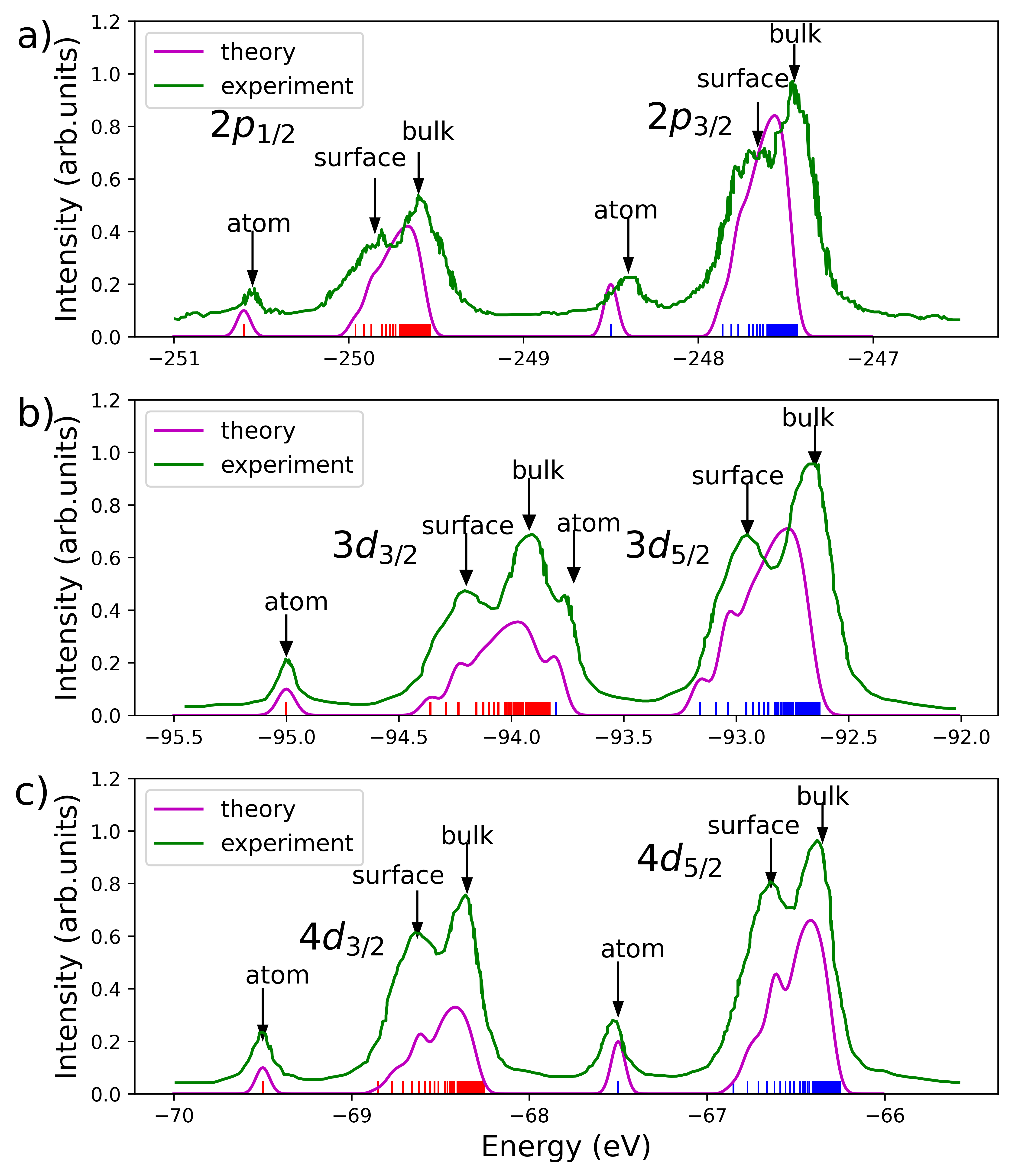}
\caption{Simulated and experimental XPS spectra of a) Ar, b) Kr, and c) Xe cluster of around 3000 atoms for Ar and Kr; 3500 atoms for Xe.
Short vertical lines represent binding energy from each atom.
A Gaussian broadening of $50$~meV was used.
}
\label{fig:spectra_Ar}
\end{figure}

Figure~\ref{fig:spectra_Ar} shows that our simulated XPS spectra are in very good agreement with experiments for Ar, Kr, and Xe clusters \cite{Tchaplyguine}.
The peaks of an isolated noble gas atom are also added in simulated spectra to indicate the relative position with respect to the two partly-resolved peaks (for each spin-orbit component) at lower  and higher BEs, originally ascribed to the XPS signal of bulk and surface atoms \cite{Tchaplyguine}.
It emerges that the experimental 'surface' peak consists of several contributions from atoms at and close to the surface, setting a range of low polarization energy. 
The experimental 'bulk' peak is from atoms deeper in the cluster. 
However, the atom close to the sphere center weakly contributes to the spectra because of their little number, and since the corresponding photoelectron can hardly reach the detector without undergoing secondary scattering events.

\paragraph*{Conclusion. ---}
Embedded many-body perturbation theory calculations on noble atom clusters of nanometric size have been presented as an ideal illustrative example of the major and often overlooked role played by electronic polarization on core energy levels.
The polarization energy is found to be orbital-independent for core levels, but highly sensitive to the position of the target atom in the cluster.
An analytical formula derived from continuum classical electrostatics captures well the position dependence of the polarization energy, thus allowing to simulate of XPS spectra with only a single $GW$ calculation.
Simulated spectra closely reproduce experiments, shedding light on the nature of the otherwise ambiguous nature of the 'surface' and 'bulk' peaks of noble gases nanoclusters. 

In general, we expect this work to raise awareness on the importance of dynamic electronic polarization effects in core-level spectroscopies. 
These are crucial for obtaining the absolute values for the binding energy of core levels in a condensed phase, but also to capture the dependence of the deep states energies of a given element on its local atomic environment.

\begin{acknowledgments}
\paragraph{Acknowledgments.---}
Part of the calculations were using the allocation of computational resources from GENCI–IDRIS (Grant 2023-A0130912036).
G.D. acknowledges support from the French "Agence Nationale de la Recherche", project RAPTORS (ANR-21-CE24-0004-01).
For the purpose of Open Access, a CC-BY public copyright licence has been applied by the authors to the present document and will be applied to all subsequent versions up to the Author Accepted Manuscript arising from this submission. 
\end{acknowledgments}

\bibliography{main}
\end{document}